\documentclass[letterpaper, 10 pt, conference]{ieeeconf}  

\IEEEoverridecommandlockouts                              
\usepackage{amsmath} 
\usepackage{amsfonts}

\newtheorem{proposition}{Proposition}[section]
\newtheorem{exmp}{Example}[section]
\usepackage{float} 
\usepackage[skip=1pt,font=footnotesize]{caption}

\usepackage{tikz}
\usetikzlibrary{shapes,arrows,calc,positioning}
\tikzstyle{bigblock} = [draw, fill=blue!20, rectangle, 
    minimum height=6em, minimum width=8em]
\tikzstyle{medblock} = [draw, fill=blue!20, rectangle, 
    minimum height=4em, minimum width=4em]    
\tikzstyle{mux} = [draw, fill=black!20, rectangle, 
    minimum height=5em, minimum width=0.1em]    
\tikzstyle{smallblock} = [draw, fill=blue!20, rectangle, 
    minimum height=2em, minimum width=3em]
    
\tikzstyle{data_block} = [draw, fill=green!20, rectangle, 
    minimum height=2em, minimum width=3em]
\tikzstyle{ops_block} = [draw, fill=blue!20, rectangle, 
    minimum height=2em, minimum width=3em]    
\tikzstyle{est_block} = [draw, fill=red!20, rectangle, 
    minimum height=2em, minimum width=3em]    
    
\tikzstyle{sum} = [draw, fill=blue!20, circle, node distance=1cm,minimum height=0.5cm]
\tikzstyle{signal} = [coordinate]
\tikzstyle{pinstyle} = [pin edge={to-,thin,black}]
\tikzstyle{block} = [draw, fill=blue!20, rectangle, 
    minimum height=3em, minimum width=9em]
\tikzstyle{blockS} = [draw, fill=blue!20, rectangle, 
    minimum height=3em, minimum width=4em]    
\tikzstyle{input} = [coordinate]
\tikzstyle{output} = [coordinate]
\usetikzlibrary{matrix}

\usetikzlibrary{positioning}
\usetikzlibrary{math}

\usepackage{hyperref}
\usepackage{xcolor}
\hypersetup{
    colorlinks,
    linkcolor={blue!100!black},
    citecolor={blue!50!black},
    urlcolor={blue!80!black}
}

\newcommand{\bc}{\begin{center}}
\newcommand{\ec}{\end{center}}
\newcommand{\benum}{\begin{enumerate}}
\newcommand{\eenum}{\end{enumerate}}
\newcommand{\nn}{\nonumber}
\newcommand{\matl}{\left[ \begin{array}}
\newcommand{\matr}{\end{array} \right]}

\renewcommand{\matl}{\begin{bmatrix}}
\renewcommand{\matr}{\end{bmatrix} }

\newcommand{\matls}{\left[ \begin{smallmatrix}}
\newcommand{\matrs}{\end{smallmatrix} \right]}
\newcommand{\isdef}{\stackrel{\triangle}{=}}

\newcommand{\tr}{{\rm tr}\,}

\newcommand{\rmT}{{\rm T}}

\newcommand{\rmd}{{\rm d}}

\newcommand{\rms}{{\rm s}}

\newcommand{\BBR}{{\mathbb R}}

\newcommand{\SN}{{\mathcal N}}

\usepackage[style=ieee ,sorting=none,maxbibnames=99,giveninits, doi=false]{biblatex}
\usepackage{xcolor,colortbl}

\title{\LARGE \bf
Modified Unscented Kalman Filter
}

\title{\LARGE \bf
Two Modifications of the Unscented Kalman Filter that\\ Specialize to the Kalman Filter for Linear Systems
}

\title{\LARGE \bf
On the Accuracy of the One-step UKF and the Two-step UKF 
}

\author{Ankit Goel and Dennis S. Bernstein
\thanks{Ankit Goel is an Assistant Professor in the Department of Mechanical Engineering, University of Maryland, Baltimore County, MD  21250. {\tt\small ankgoel@umbc.edu}.}%
\thanks{Dennis S. Bernstein is a Professor in the Department of Aerospace Engineering, University of Michigan, Ann Arbor, MI 48109. {\tt\small dsbaero@umich.edu}.}%
}

\begin{document}

\maketitle
\thispagestyle{empty}
\pagestyle{empty}


\begin{abstract}
    %
    The most accurate version of the unscented Kalman filter (UKF) involves the construction of two ensembles.
    To reduce computational cost, however, UKF is often implemented without the second ensemble. 
    This simplification comes at a price, however, since, for linear systems, the one-step variation of the two-step UKF does not specialize to the classical Kalman filter, with an associated loss of accuracy.
    This paper remedies this drawback by developing a modified one-step UKF that recovers the classical Kalman filter for linear systems.
    %
    %
    %
    Numerical examples show that the modified one-step UKF also recovers the accuracy of the two-step UKF in nonlinear systems with linear outputs. 
    %
\end{abstract}




\section{Introduction}

The Unscented Kalman filter (UKF) is widely applied to  nonlinear estimation problems \cite{simon2006optimal}.
UKF was introduced in \cite{wan2000unscented,van2001square} has been applied to a wide array of engineering and scientific applications including attitude estimation  \cite{kraft2003quaternion}, 
navigation  \cite{gao2018multi}, 
battery-charge estimation \cite{he2016real}, and
state and parameter estimation in atmospheric models \cite{gove2006application}.

Like the Ensemble Kalman filter (EnKF) \cite{anderson2001ensemble}, UKF propagates an ensemble in order to compute the mean and covariance of the state estimate. 
However, unlike EnKF, which approximates the covariance using statistics of the propagated ensembles, UKF uses unscented transformations to approximate the covariances, which allows UKF to reduce the size of the ensemble to $2n+1,$ where $n$ is the dimension of the state of the system \cite{julier1997new}.   
Since UKF propagates the ensemble using the nonlinear dynamics map, the accuracy of UKF is expected and is also reported to be better than that of the Extended Kalman filter, which is based on the linearized dynamics \cite{st2004comparison}.

The classical UKF requires generation of a $2n+1$ size ensemble twice at every step\cite[p.~447]{simon2006optimal}, \cite[p.~86]{sarkka2013bayesian}.
The first ensemble is used to propagate the estimated state and compute the prior covariance, whereas the second ensemble is used to approximate cross-covariance matrices needed to compute the filter gain. 
This is the \textit{two-step UKF}.
%
Since the UKF gain and covariance update are motivated by the corresponding expressions used in the Kalman filter, it is reasonable to expect that, in the case of a linear system, the UKF gain and the covariance update will coincide with Kalman filter.
As expected, the two-step UKF indeed specializes to the classical Kalman filter when applied to a linear system, 
as explicitly shown in Section \ref{sec:summaryUKF}.


In the two-step UKF, the prior estimate and the prior covariance computed after propagating the first ensemble through the dynamics of the system are used to generate the second ensemble, which is then further transformed into the output ensemble using the algebraic output equation. 
In order to reduce implementation complexity and reduce computational cost, second ensemble generation is often omitted.
Instead, the propagated ensemble is used for further computations. 
This is the \textit{one-step UKF}. 
In fact, the UKF originally introduced in \cite{julier1997new,wan2000unscented,van2001square} presented the one-step formulation. 
However, it turns out that, the one-step UKF does not specialize to the classical Kalman filter when applied to a linear system. 
This is due to the fact that effect of the process noise does not pass through to the output-error covariance.
In fact, one-step UKF output covariances and the propagated state covariance are found to be missing the process noise term when applied to a linear system, as shown in this paper.

This paper presents a modification of the one-step UKF that recovers the accuracy of the two-step UKF for systems where the output equation is linear with only one ensemble generation.
Like the two-step UKF, the one-step modified UKF (MUKF) specializes to the Kalman filter for linear systems. 
In particular, we show explicitly that the two-step UKF specializes to the the Kalman filter in the case of a linear system.
Next, we show that the accuracy of the one-step UKF is worse than the accuracy of the two-step UKF in the case of linear system by explicitly stating the missing terms.
Finally, by including the missing terms, we present the one-step modified UKF that recovers the accuracy of the two-step UKF.

This paper is organized as follows.
Section \ref{sec:summaryKF} briefly reviews the Kalman filter to introduce the terminology and notation used in this paper. 
Section \ref{sec:summaryUKF} briefly reviews the two-step UKF and shows that, for linear systems, it specializes to the classical Kalman filter.
Section \ref{sec:one_step_UKF} reviews the one-step UKF to a linear system and shows that, for linear systems, it does not specialize to the classical Kalman filter.
Section \ref{sec:EUKF} presents a modification of the one-step UKF that specializes to the classical Kalman filter in the case of linear systems. 
Section \ref{sec:exmple} applies the proposed extension to two nonlinear systems and compares the accuracy of uncertainty propagation.
Finally, the paper concludes with a discussion in Section \ref{sec:conclusion}.

\section{Summary of the Kalman Filter}
\label{sec:summaryKF}
This section briefly reviews the Kalman filter to introduce terminology and notation for later sections.
Consider a linear system
\begin{align}
    x_{k+1}
        &= 
            A_k x_{k} + B_k u_{k} + w_{k },
    \label{eq:LS_xkp1}
    \\
    y_{k} 
        &=
            C_k x_{k} + v_{k },
    \label{eq:LS_yk}
\end{align}
where, for all $k \ge 0$, 
$A_k, B_k, C_k$ are real matrices,
$w_k \sim \SN(0,Q_k)$ is the disturbance, and 
$v_k \sim \SN(0,R_k)$ is the sensor noise.  

For the system \eqref{eq:LS_xkp1}, \eqref{eq:LS_yk}, the Kalman filter is 
%
%
\begin{align}
    \hat x_{k+1|k}
        &=
            A_k \hat x_{k|k} + B_k u_k,
    \label{eq:KF_x(k+1|k)}
        \\
    \hat x_{k+1|k+1}
        &=
            \hat x_{k+1|k} +
            K_{k+1}
            (y_{k+1} - C_{k+1} \hat x_{k+1|k}),
    \label{eq:KF_x(k+1|k+1)}
\end{align}
where 
$\hat x_{k+1|k}$ is the \textit{prior estimate},
$\hat x_{k+1|k+1}$ is the \textit{posterior estimate} at step $k+1,$ and 
the Kalman gain $K_{k+1 } $ is given by
\begin{align}
    K_{k+1 } 
        &=
            P_{k+1|k} C_{k+1}^\rmT
            \overline R_{k+1}^{-1}.
    \label{eq:2SOF_minimizer}
\end{align}
The prior covariance $P_{k+1|k}$ at step $k+1$ is given by
\begin{align}
    P_{k+1|k}
        &=
            A_k P_{k|k} A_k^\rmT + Q_k,
    \label{eq:2SOF_prior_cov}
\end{align}
and the posterior covariance at step $k+1$ is given by
    \begin{align}
        &P_{k+1|k+1}
            =
                P_{k+1|k} 
                -
                P_{k+1|k} C_{k+1}^\rmT
                \overline R_{k+1}^{-1}
                C_{k+1} P_{k+1|k} .
           \label{eq:2SOF_optimized_post_cov}
    \end{align}
where 
\begin{align}
    \overline R_{k+1}
        \isdef 
            C_{k+1} P_{k+1|k} C_{k+1}^\rmT + R_{k+1}.
\end{align}
The \textit{Kalman filter} is \eqref{eq:KF_x(k+1|k)}, \eqref{eq:KF_x(k+1|k+1)} where $ K_{k+1}$ is given by 
\eqref{eq:2SOF_minimizer}
and the covariance matrices are updated by \eqref{eq:2SOF_prior_cov}, \eqref{eq:2SOF_optimized_post_cov}.

Next, in order to establish connections with UKF, 
\eqref{eq:2SOF_minimizer}, \eqref{eq:2SOF_optimized_post_cov} 
are reformulated in terms of covariance matrices instead of $A_k$ and $C_{k+1}$. 
%
%
Defining 
\begin{align}
    P_{z_{k+1|k+1}}
        &\isdef
            C_{k+1} P_{k+1|k} C_{k+1}^\rmT + R_{k+1}, 
    \label{eq:2SOF_Pz}
        \\
    P_{e,z_{k+1|k}}
        &\isdef
            P_{k+1|k} C_{k+1}^\rmT,
    \label{eq:2SOF_Pez}
\end{align}
%
%
and
substituting 
\eqref{eq:2SOF_Pz} and
\eqref{eq:2SOF_Pez} in \eqref{eq:2SOF_minimizer} and
\eqref{eq:2SOF_optimized_post_cov},
the Kalman gain can be written as
\begin{align}
    K_{k+1 } 
            &=
                P_{e,z_{k+1|k}} 
                P_{z_{k+1|k+1}} ^{-1} ,
    \label{eq:KF_opt_gain_cov}
\end{align}
and the corresponding optimized posterior covariance at step $k+1$ can be written as
\begin{align}
    P_{k+1|k+1}
        &=
            P_{k+1|k} -
            K_{k+1 } 
            P_{e,z_{k+1|k}} ^\rmT
            .
    \label{eq:KF_opt_cov_cov}
\end{align}
As shown in the next section, UKF approximates the covariance matrices $P_{k+1|k},$ $P_{z_{k+1|k+1}},$ and $P_{e,z_{k+1|k}}$ by using ensembles instead of $A_k$ and $C_{k+1}.$

\section{Summary of Two-Step UKF}
\label{sec:summaryUKF}

This section briefly reviews the classical two-step unscented Kalman filter to establish notation and terminology for use in the rest of the paper. 
The UKF algorithm is formulated using a compact matrix-based notation and is based on the algorithm presented in  
\cite[p.~86]{sarkka2013bayesian}.

Consider a system
\begin{align}
    x_{k+1}
        &= 
            f_k (x_{k}, u_{k}) + w_{k },
    \label{eq:NLS_xkp1}
    \\
    y_{k} 
        &=
            g_k( x_{k}) + v_{k },
    \label{eq:NLS_yk}
\end{align}
where, for all $k \ge 0$, 
$f_k, g_k, C_k$ are real-valued vector functions,
$w_k \sim \SN(0,Q_k)$ is the disturbance, and 
$v_k \sim \SN(0,R_k)$ is the sensor noise.

The following notation is used to present ensembles in a compact manner. 
Let $x \in \BBR^{l_x}$ and $P \in \BBR^{l_x \times l_x}$ be positive definite. 
The ensemble $X(x,P) \in \BBR^{l_x \times (2 l_x+1)}$ is the matrix
\begin{align}
    X(x,P) 
        \isdef 
            [
                x  \
                &
                x + p_1 \
                \cdots \
                x + p_{l_x} \
                x - p_1 \
                \cdots \
                x - p_{l_x} 
            ],\nn 
\end{align}
where $p_i$ is the $i$-th column of $P.$
Let $\alpha>0.$ 
Define
\begin{align}
    W
        \isdef
            \dfrac{1}{2\alpha^2 l_x}
            \matl
                2 (\alpha^2 -1 ) l_x \\ 
                 1_{2 l_x \times 1} \\
            \matr\in\BBR^{2l_x+1}. \nn 
\end{align}
The weighted mean of the ensemble $X$ is $\bar x \isdef X W,$ and the ensemble perturbation is
$\tilde X 
        \isdef 
            X - H(\bar x),
$
where, for $ v \in \BBR^n,$
$
H(v) 
        \isdef
            1_{1 \times 2 l_x+1 } \otimes v
            \in \BBR^{n  \times (2 l_x+1) }
            .
$
Note that $\otimes$ is the Kronecker product \cite{van2000ubiquitous}.


In order to compute the filter gain $K_{k+1}$
and the posterior covariance $P_{k+1|k+1},$
UKF approximates the covariance matrices  
$P_{k+1|k},$
$P_{z_{k+1|k+1}},$ and
$P_{e,z_{k+1|k}}$ in \eqref{eq:KF_opt_gain_cov} and \eqref{eq:KF_opt_cov_cov} by propagating an ensemble of $2l_x+1$ sigma points. 

For all $k \ge 0$, the $i$-th sigma point $\hat x_{\sigma_i,k}$ is defined as the $i$-th column of 
\begin{align}
    X_{k|k}
        \isdef 
            X
            \left(
                \hat x_{k|k}, \alpha  \sqrt{ l_x  P_{k|k}}
            \right)
        ,
    \label{eq:Xk_sigma_matrix}
\end{align}
where $\alpha\in \BBR$ is a tuning parameter and $P_{k|k}$ is the posterior covariance given by UKF at step $k$.
Then, for $i = 1, \ldots, 2 l_x+1,$ the sigma points are propagated as 
\begin{align}
    \hat x_{\sigma_i,k+1} 
        &=
            f_k(\hat x_{\sigma_i,k}, u_k).
    \label{eq:x_sigma_propagated_FS}        
\end{align}
The prior estimate and the prior covariance at step $k+1$ are given by 
\begin{align}
    \hat x_{k+1|k}
        &=
            X_{k+1|k} W,
    \label{eq:UKF_prior_x}
    \\
    P_{k+1|k}
        &=
            \tilde X_{k+1|k} W_\rmd  \tilde X_{k+1|k}^\rmT  + Q_k,
    \label{eq:UKF_prior_P}
\end{align}
where 
\begin{align}
    X_{k+1|k}
        &\isdef
            \matl
                \hat x_{\sigma_1,k+1}   & \cdots    & \hat x_{\sigma_{2l_x+1},k+1} 
            \matr.
    \label{eq:propagated_ensemble_UKF}
\end{align}

Next, the posterior estimate and the posterior covariance at step $k+1$ are computed by regenerating sigma points as shown next. 
Defining 
\begin{align}
    X_{k+1|k}'
        \isdef 
            X
            \left(
            \hat x_{k+1|k}, \alpha  \sqrt{ l_x P_{k+1|k}}
            \right),
    \label{eq:Xk_prior_sigma_matrix}
\end{align}
the output of the $i$-th sigma point is given by
\begin{align}
    \hat y_{\sigma_i,k+1} 
        &=
            g_{k+1}(X_{k+1|k}' e_i),
    \label{eq:y_sigma_propagated_FS}    
\end{align}
where $e_i$ is the $i$-th column of $I_{2 l_x+1}.$
The covariance matrices $P_{z_{k+1|k+1}}$ and $P_{e,z{k+1|k}}$ are then given by
\begin{align}
    P_{z_{k+1|k+1}}
        &=
            \tilde Y_{k+1} W_\rmd \tilde Y_{k+1}^\rmT  + R_{k+1},
    \label{eq:Pz_UKF_FS}
        \\
    P_{e,z_{k+1|k}}
        &=
            \tilde X_{k+1|k}' W_\rmd \tilde Y_{k+1}^\rmT,
    \label{eq:Pez_UKF_FS}
\end{align}
where 
\begin{align}
    Y_{k+1}
        \isdef
            \matl
                \hat y_{\sigma_1,k+1}   & \cdots    & \hat y_{\sigma_{2l_x+1},k+1} 
            \matr
            \in \BBR^{l_y \times 2 l_x+1}.
    \label{eq:y_ensemble}
\end{align}
Finally, the posterior estimate at step $k+1$ is 
\begin{align}
    \hat x_{k+1|k+1}
        &=
            \hat x_{k+1|k} +
            K_{k+1 } 
            (y_{k+1} - Y_{k+1} W),
        \label{eq:UKF_post_update}
\end{align}
and the posterior covariance at step $k+1$ is 
\begin{align}
    P_{k+1|k+1}
        &=
        P_{k+1|k} 
        -
        K_{k+1 } 
        P_{e,z_{k+1|k}}^{\rmT}
        ,
    \label{eq:UKF_P_post_PzPez}
\end{align}
where
\begin{align}
    K_{k+1 } 
        &=
            P_{e,z_{k+1|k}} 
            P_{z_{k+1|k+1}}^{-1} .
    \label{eq:UKF_opt_PzPez}
\end{align}
The \textit{two-step UKF} is \eqref{eq:UKF_prior_x}, \eqref{eq:UKF_post_update} where 
the posterior covariance is given by \eqref{eq:UKF_P_post_PzPez} and 
the filter gain is given by 
\eqref{eq:UKF_opt_PzPez}.
Note that \eqref{eq:UKF_P_post_PzPez}, \eqref{eq:UKF_opt_PzPez} are similar to and are in fact motivated by \eqref{eq:KF_opt_gain_cov}, \eqref{eq:KF_opt_cov_cov}.
The various covariance matrices computed in the two-step UKF are summarized in Table \ref{tab:UKF_mods}.

\renewcommand{\arraystretch}{1.2}
\begin{table*}[h]
    \normalsize 
    \centering
    \begin{tabular}{|c|c|c|c|}
        \hline
        Variable & Two-step UKF & One-step UKF & Modified One-step UKF \\
        \hline
            $X_{k|k}$ &
            $X(\hat x_{k|k}, \alpha \sqrt{P_{k|k}})$ &
            $X(\hat x_{k|k}, \alpha \sqrt{P_{k|k}})$ &
            $X(\hat x_{k|k}, \alpha \sqrt{P_{k|k}})$
            \\
            & \eqref{eq:Xk_sigma_matrix}
            & \eqref{eq:Xk_sigma_matrix}
            & \eqref{eq:Xk_sigma_matrix}
        \\
        \hline
            $P_{k+1|k}$ &
            $\tilde X_{k+1|k} W_\rmd  \tilde X_{k+1|k}^\rmT  + Q_k$ &
            $\tilde X_{k+1|k} W_\rmd  \tilde X_{k+1|k}^\rmT  + Q_k$ &
            $\tilde X_{k+1|k} W_\rmd  \tilde X_{k+1|k}^\rmT  + Q_k$
            \\
        & \eqref{eq:UKF_prior_P} & \eqref{eq:UKF_prior_P} & \eqref{eq:UKF_prior_P}
            \\
        \hline
            $X_{k+1|k}'$ &
            \color{black}{$X(\hat x_{k+1|k}, \alpha \sqrt{P_{k+1|k}})$} &
            $X_{k+1|k}$ &
            $X_{k+1|k}$
            \\
        & \eqref{eq:Xk_prior_sigma_matrix} 
        & \eqref{eq:propagated_ensemble_UKF}
        & \eqref{eq:propagated_ensemble_UKF}
        \\
        \hline
            $P_{z_{k+1|k+1}}$ &  
            $\tilde Y_{k+1} W_\rmd  \tilde Y_{k+1}^\rmT$ &
            $\tilde Y_{k+1} W_\rmd  \tilde Y_{k+1}^\rmT$ &
            \color{black}{$\tilde Y_{k+1} W_\rmd  \tilde Y_{k+1}^\rmT + Q_k C_{k+1}^\rmT$} 
            \\
        & \eqref{eq:Pz_UKF_FS} & \eqref{eq:Pz_UKF_1step} & \eqref{eq:Pz_MUKF_1step}
            \\
        \hline
            $P_{ez,_{k+1|k}}$ &
            $\tilde X_{k+1|k}' W_\rmd  \tilde Y_{k+1}^\rmT  + R_{k+1}$ &
            $\tilde X_{k+1|k} W_\rmd  \tilde Y_{k+1}^\rmT  + R_{k+1}$ &       
            \color{black}{$\tilde X_{k+1|k} W_\rmd  \tilde Y_{k+1}^\rmT  + C_{k+1} Q_k C_{k+1}^\rmT + R_{k+1}$ }
            \\
        & \eqref{eq:Pez_UKF_FS} & \eqref{eq:Pez_UKF_1step} & \eqref{eq:Pez_MUKF_1step}
            \\
        \hline
    \end{tabular}
    \caption{Ensembles and covariance matrices used in the two-step UKF, the one-step UKF, and the modified one-step UKF.}
    \label{tab:UKF_mods}
\end{table*}

The following result shows that the two-step UKF specializes to the Kalman filter when applied to a linear system.

\begin{proposition}
    \label{prop:apply_UKF_to_LS}
    Consider the linear system \eqref{eq:LS_xkp1}, \eqref{eq:LS_yk}.
    %
    %
    Let $P_{k|k}$ be the posterior covariance given by the Kalman filter
    and 
    let $P_{k|k}^{\rm UKF}$ be the posterior covariance given by the two-step UKF.
    Let $\hat x \in \BBR^n$ and let $P$ be positive definite. 
    Assume that 
    $
    \hat x_{k|k}^{\rm UKF}
            =
                \hat x_{k|k}^{\rm KF}
            =
                \hat x
    $ and $
        P_{k|k}^{\rm UKF}
            =
                P_{k|k}^{\rm KF} 
            = 
                P.
    $
    Then,
    \begin{align}
        P_{z_{k+1|k+1}}^{UKF}
            &=
                P_{e,z_{k+1|k}}^{KF}, 
        \\
        P_{e,z_{k+1|k}}^{UKF}
            &=
                P_{e,z_{k+1|k}}^{KF}. 
    \end{align}
    Furthermore, 
    \begin{align}
        \hat x_{k+1|k+1}^{\rm UKF}
            &=
                \hat x_{k+1|k+1}^{\rm KF},
        \label{eq:UKF2_x_eq_KF_x}
        \\
        K_{k+1}^{\rm UKF}
            &=
                K_{k+1},
        \label{eq:UKF2_K_eq_KF_K}
        \\
        P_{k+1|k+1}^{\rm UKF}
            &=
                P_{k+1|k+1}^{\rm KF}.
        \label{eq:UKF2P_eq_KFP}
    \end{align}
\end{proposition}
\begin{proof}
See Appendix \ref{proof:2UKF}.
\end{proof}

Proposition \ref{prop:apply_UKF_to_LS} implies that, in a linear system, the two-step UKF reduces to the Kalman filter.
%
Furthermore, note that, in linear systems, the choice of $\alpha$ does not affect $K_{k+1}^{\rm UKF}$ and $P_{k+1|k+1}^{\rm UKF}.$ 

\section{One-Step UKF}
\label{sec:one_step_UKF}
This section reviews the one-step UKF presented in \cite{julier1997new,wan2000unscented,van2001square,simon2006optimal}, where the second ensemble generation step, given by \eqref{eq:Xk_prior_sigma_matrix}, is omitted in order to reduce computational effort and cost.


In this case, the output of the $i$-th sigma point is given by
\begin{align}
    \hat y_{\sigma_i,k+1} 
        &=
            g_{k+1}(X_{k+1|k} e_i),
    \label{eq:y_sigma_propagated_UKF1}    
\end{align}
which uses the propagated ensemble $X_{k+1|k}$ given by \eqref{eq:propagated_ensemble_UKF}, instead of regenerating a new ensemble using the prior estimate and the prior covariance.  
In the one-step UKF, the covariance matrices $P_{z_{k+1|k+1}}$ and $P_{e,z{k+1|k}}$ are given by
\begin{align}
    P_{z_{k+1|k+1}}
        &=
            \tilde Y_{k+1} W_\rmd \tilde Y_{k+1}^\rmT  + R_{k+1},
    \label{eq:Pz_UKF_1step}
        \\
    P_{e,z_{k+1|k}}
        &=
            \tilde X_{k+1|k} W_\rmd \tilde Y_{k+1}^\rmT.
    \label{eq:Pez_UKF_1step}
\end{align}
Note that \eqref{eq:Pz_UKF_1step} and \eqref{eq:Pez_UKF_1step} use the propagated ensemble $X_{k+1|k}$ to compute the perturbed ensembles instead of using the regenerated ensemble $X_{k+1|k}'.$

The \textit{ one-step UKF} is \eqref{eq:UKF_prior_x}, \eqref{eq:UKF_post_update} where 
the posterior covariance is given by \eqref{eq:UKF_P_post_PzPez} and 
the filter gain is given by 
\eqref{eq:UKF_opt_PzPez},
However $P_{z_{k+1|k+1}}$ and $P_{e,z_{k+1|k}}$ used in \eqref{eq:UKF_P_post_PzPez}, \eqref{eq:UKF_opt_PzPez} are now given by \eqref{eq:Pz_UKF_1step}, \eqref{eq:Pez_UKF_1step}.
The various covariance matrices computed in the one-step UKF are summarized in Table \ref{tab:UKF_mods}.

The following result shows that the  one-step UKF does not specialize to the Kalman filter when applied to a linear system.

\begin{proposition}
    \label{prop:apply_1UKF_to_LS}
    Consider the linear system \eqref{eq:LS_xkp1}, \eqref{eq:LS_yk}.
    Let $P_{k|k}$ be the posterior covariance given by the Kalman filter
    and 
    let $P_{k|k}^{\rm UKF1}$ be the posterior covariance given by the  one-step UKF.
    Let $\hat x \in \BBR^n$ and let $P$ be positive definite. 
    Assume that 
    $
    \hat x_{k|k}^{\rm UKF1}
            =
                \hat x_{k|k}^{\rm KF}
            =
                \hat x
    $ and $
        P_{k|k}^{\rm UKF1}
            =
                P_{k|k}^{\rm KF} 
            = 
                P.
    $
    Then,
    \begin{align}
        P_{z_{k+1|k+1}}^{UKF1}
            &=
                P_{e,z_{k+1|k}}^{KF} - C_{k+1} Q_k C_{k+1}^\rmT, 
        \\
        P_{e,z_{k+1|k}}^{UKF1}
            &=
                P_{e,z_{k+1|k}}^{KF} - Q_k C_{k+1}^\rmT, 
    \end{align}
    Furthermore, assume that 
    $Q_k \neq 0,$ and $C_k \notin \SN(Q_k).$
    Then, 
    \begin{align}
        \hat x_{k+1|k+1}^{\rm UKF1}
            &\neq
                \hat x_{k+1|k+1}^{\rm KF},
        \label{eq:UKF1_x_neq_KF_x}
        \\
        K_{k+1}^{\rm UKF1}
            &\neq
                K_{k+1},
        \label{eq:UKF1_K_neq_KF_K}
        \\
        P_{k+1|k+1}^{\rm UKF1}
            &\neq
                P_{k+1|k+1}^{\rm KF},
        \label{eq:UKF1P_neq_KFP}
        \\
        \tr (P_{k+1|k+1}^{\rm UKF1})
            &\le
                \tr ( P_{k+1|k+1}^{\rm KF}).
        \label{eq:UKF1P_le_KFP}
    \end{align}

\end{proposition}
\begin{proof}
See Appendix \ref{proof:1UKF}.
\end{proof}

Proposition \ref{prop:apply_1UKF_to_LS} implies that, in a linear system  where disturbance is not zero, the one-step UKF does not reduce to the Kalman filter.
That is, the posterior covariance propagated by the one-step UKF is not equal to the covariance given by \eqref{eq:2SOF_optimized_post_cov}.
%
This inequality arises due to the fact that the output error covariance $P_{z_{k+1|k+1}}$ is missing the term $C_{k+1} Q_k C_{k+1}^\rmT $ and the cross-covariance $P_{e,z_{k+1|k}}$ is missing the term $Q_k C_{k+1}^\rmT.$
The next section presents a modification of the one-step UKF that includes the missing term and is thus more accurate than the one-step UKF.
This modification is expecially beneficial for high-dimension nonlinear systems where the second sigma-point generation step adds considerable computational cost. 
Furthermore, the second sigma-point generation step makes the algorithm non-modular. 

%

 
%
\section{One-Step Modified UKF}
\label{sec:EUKF}

As shown in the previous section, the covariances $P_{z_{k+1|k+1}}$ and $P_{e,z_{k+1|k}}$ in \eqref{eq:UKF_Pz} and \eqref{eq:UKF_Pez}  are missing terms that depend on the disturbance statistics $Q_k$, thus preventing one-step UKF from specializing to the Kalman filter for linear systems.
To remedy this omission, this section presents the one-step modified UKF (MUKF), which specialize to the Kalman filter for linear systems.
In this modification, the UKF covariance matrices \eqref{eq:Pz_UKF_1step}, \eqref{eq:Pez_UKF_1step} are modified such that they specialize to \eqref{eq:2SOF_Pz},  \eqref{eq:2SOF_Pez} in the case of linear systems.

Using the output matrix, MUKF adds the missing terms to $P_{z_{k+1|k+1}}$ and $P_{e,z_{k+1|k}}$.
In particular, in MUKF, the covariance matrices $P_{z_{k+1|k+1}}$ and $P_{e,z{k+1|k}}$ are given by
\begin{align}
    P_{z_{k+1|k+1}}
        &=
            \tilde Y_{k+1} W_\rmd \tilde Y_{k+1}^\rmT  
            + C_{k+1} Q_k C_{k+1}^\rmT
            + R_{k+1},
    \label{eq:Pz_MUKF_1step}
        \\
    P_{e,z_{k+1|k}}
        &=
            \tilde X_{k+1|k} W_\rmd \tilde Y_{k+1}^\rmT
            +
            Q_k C_{k+1}^\rmT.
    \label{eq:Pez_MUKF_1step}
\end{align}
%
Note that, in the case of nonlinear systems, $C_{k+1}$ can be computed using the Jacobian of the output map.

Since, in the case of linear systems, the intermediate covariance matrices in MUKF include the missing terms, the one-step modified UKF recovers the accuracy of the classical two-step UKF. 
The next section applies the MUKF to two nonlinear systems to demonstrate this fact.

\section{Numerical Examples}
\label{sec:exmple}
In this section,  the two-step UKF, one-step UKF, and the one-step MUKF  are applied to two nonlinear systems, namely, the Van der Pol Oscillator and the chaotic Lorenz system to demonstrate the erroneous covariance update in the one-step UKF and the recovery of the correct covariance update in the one-step MUKF. 



\begin{figure}[h]
    \centering
    \includegraphics[width = 1\columnwidth]{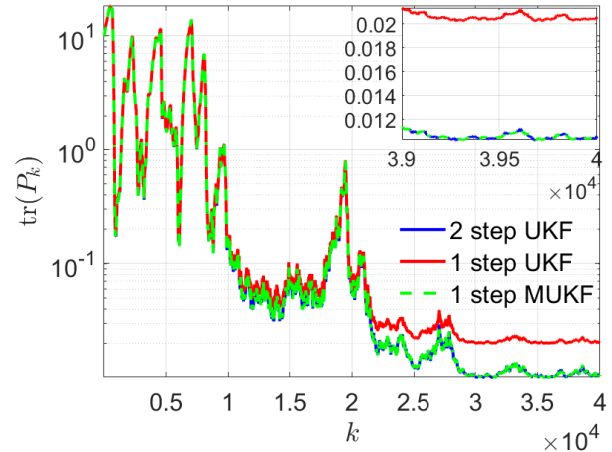}
    \caption{
    Example \ref{exmp:VDP}. Trace of the posterior covariance computed using the two-step UKF, the one-step UKF, and the one-step MUKF on a log scale with a zoomed-in inset showing the last 1000 steps of the simulation. 
    Note that the one-step MUKF recovers the accuracy of the two-step UKF. 
    }
    \label{fig:CDC_2022_MUKF_VDP_Pk}
\end{figure}
\begin{figure}[h]
    \centering
    \includegraphics[width = 1\columnwidth]{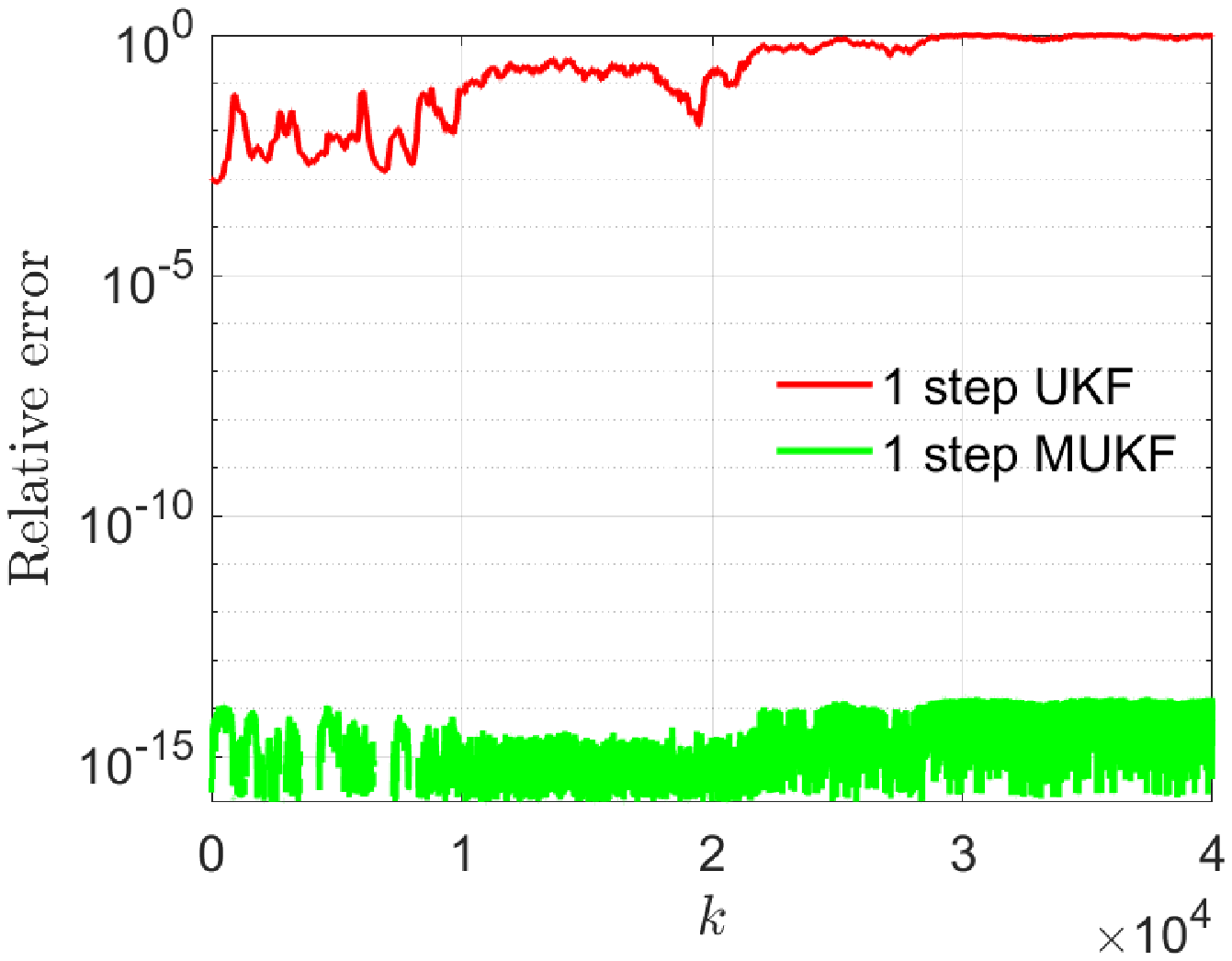}
    \caption{
    Example \ref{exmp:VDP}.
    Relative error in the trace of the posterior covariance computed using the one-step UKF, and the one-step MUKF with respect to the two-step UKF.
    Note that the posterior covariance computed by the one-step MUKF is numerically within the machine precision of the two-step UKF posterior covariance, whereas the posterior covariance computed by the one-step UKF is almost twice the two-step UKF posterior covariance.
    }
    \label{fig:CDC_2022_MUKF_VDP_rel_error}
\end{figure}

\begin{exmp}
    \label{exmp:VDP}
    \textit{Van der Pol Oscillator}.
Consider the discretized Van der Pol Oscillator. 
\begin{align}
    x_{k+1} 
        &=
            f(x_k) + w_k,
\end{align}
where 
\begin{align}
    f(x)
        =
            \matl
                x_1 + T_\rms x_2 \\
                x_2 + T_\rms (\mu (1-x_1^2) x_2-x_1)
            \matr,
\end{align}
and $\mu = 1.2.$
Let the measurement be given by
\begin{align}
    y_k
        =
            C x_k + v_k,
\end{align}
where $C \isdef [1 \ 0].$
For all $k\ge0,$ let $Q_k = 0.01 I_2$  and $R_k = 10^{-4}.$
Furthermore, let $x(0) = [1 \  1 ]^\rmT$ and $P_{0|0} = I_3$.

Letting $\alpha = 1.5$ in the two-step UKF, the one-step UKF, and the one-step MUKF, Figure \ref{fig:CDC_2022_MUKF_VDP_Pk} shows the trace of the posterior covariance computed by the three filters. 
Note that one-step UKF posterior covariance is larger than the two-step UKF posterior covariance, whereas the one-step UKF recovers the two-step posterior covariance in spite of generating only one ensemble per step. 
Figure \ref{fig:CDC_2022_MUKF_VDP_rel_error} shows the relative error of the one-step UKF and the one-step MUKF posterior covariance relative to the two-step UKF. 
Specifically, the relative error is given by the ratio
\begin{align}
    \dfrac
        {\tr P_{k|k}^\rms - \tr P_{k|k}^{\rm UKF}}
        {\tr P_{k|k}^{\rm UKF}},
\end{align}
where $\rms=\rm UKF1$ or $\rm MUKF. $  
Note that, in this particular example, the one-step UKF relative error is almost $100\%$, whereas the one-step MUKF relative error is less than the machine precision, that is, the one-step MUKF recovers the two-step UKF.
%

%
This example shows that the one-step MUKF posterior covariance estimate is more accurate than the one-step UKF posterior covariance and is numerically equal to the two-step UKF posterior covariance.
$\hfill\mbox{\Large$\diamond$}$
\end{exmp}

\begin{figure}[h]
    \centering
    \includegraphics[width = 1\columnwidth]{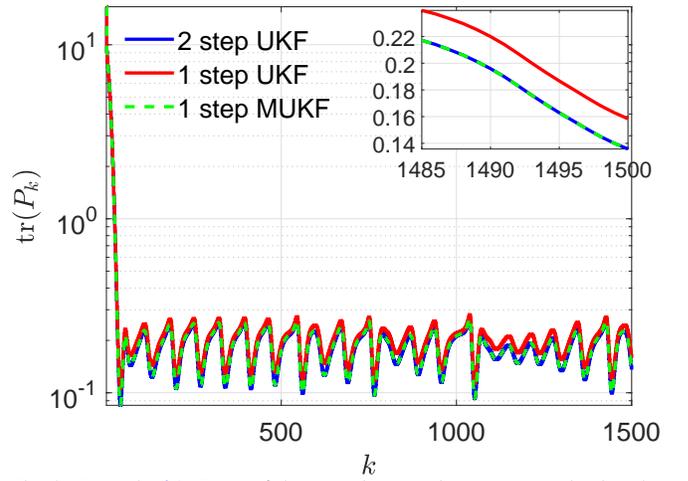}
    \caption{
    Example \ref{exmp:Lorenz}. 
    Trace of the posterior covariance computed using the two-step UKF, the one-step UKF, and the one-step MUKF on a log scale with a zoomed-in inset showing the last 15 steps of the simulation. 
    Note that the one-step MUKF recovers the accuracy of the two-step UKF. 
    }
    \label{fig:CDC_2022_MUKF_Lorenz_Pk}
\end{figure}

\begin{figure}[h]
    \centering
    \includegraphics[width = 1\columnwidth]{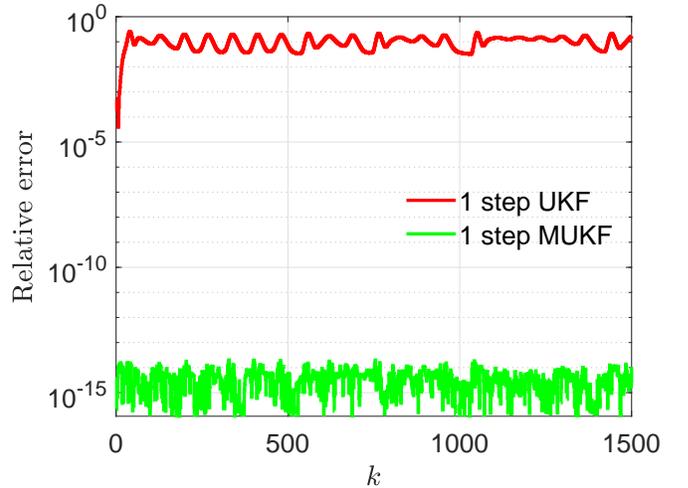}
    \caption{
    Example \ref{exmp:Lorenz}. 
    Relative error in the trace of the posterior covariance computed using the one-step UKF, and the one-step MUKF with respect to the two-step UKF.
    Note that the posterior covariance computed by the one-step MUKF is numerically within the machine precision of the two-step UKF posterior covariance, whereas the posterior covariance computed by the one-step UKF is about 15 \% larger than the two-step UKF posterior covariance.
    }
    \label{fig:CDC_2022_MUKF_Lorenz_rel_error}
\end{figure}

%
\begin{exmp}
    \label{exmp:Lorenz}
    \textit{Lorenz System}.
Consider the Lorenz system
\begin{align}
    \matl
        \dot x_1 \\
        \dot x_2 \\
        \dot x_3
    \matr
        =
            \matl
                \sigma (x_2-x_1) \\
                x_1(\rho-x_3)-x_2 \\
                x_1x_2 - \beta x_3
            \matr,
    \label{eq:Lorenz}
\end{align}
which exhibits a choatic behaviour for 
$\sigma = 10,$ $\rho = 28,$ and $\beta = 8/3$.
The Lorenz system \eqref{eq:Lorenz} is integrated using the forward Euler method with step size $T_\rms=0.01.$
Let the discrete system be modeled as
\begin{align}
    x_{k+1} 
        =
            f(x_k) + w_k,
\end{align}
where 
\begin{align}
    f(x) 
        \isdef 
            x+T_\rms 
            \matl
                \sigma (x_2-x_1) \\
                x_1(\rho-x_3)-x_2 \\
                x_1x_2 - \beta x_3
            \matr
    \label{eq:Lorenz_disc}
\end{align}
and $w_k \sim \SN(0,Q_k).$
For all $k \ge 0$, let 
\begin{align}
    y_k 
        =
            C x_k + v_k,
    \label{eq:Lorenz_output}
\end{align} 
where
$C \isdef [0 \ 1 \ 0]$ and 
$v_{k} \sim \SN(0,R_k).$
For all $k\ge0,$ let $Q_k = 0.01 I_2$  and $R_k = 10^{-4}.$
Furthermore, let $x(0) = [1 \ 1 \ 1 ]^\rmT$ and $P_{0|0} = I_3$.
%

Letting $\alpha = 1.5$ in the two-step UKF, the one-step UKF, and the one-step MUKF, Figure \ref{fig:CDC_2022_MUKF_Lorenz_Pk} shows the trace of the posterior covariance computed by the three filters. 
Note that one-step UKF posterior covariance is larger than the two-step UKF posterior covariance, whereas the one-step UKF recovers the two-step posterior covariance in spite of generating only one ensemble per step. 
Figure \ref{fig:CDC_2022_MUKF_Lorenz_rel_error} shows the relative error of the one-step UKF and the one-step MUKF posterior covariance relative to the two-step UKF. 
Note that, in this particular example, the one-step UKF relative error is almost $15\%$, whereas the one-step MUKF relative error is less than the machine precision, that is, the one-step MUKF recovers the two-step UKF.

This example shows that the one-step MUKF posterior covariance estimate is more accurate than the one-step UKF posterior covariance and is numerically equal to the two-step UKF posterior covariance.
$\hfill\mbox{\Large$\diamond$}$
\end{exmp}

\section{Conclusions}
\label{sec:conclusion}

This paper explicitly showed that the two-step UKF specialize to the classical Kalman filter for linear systems, whereas the one-step UKF does not. 
Consequently, the accuracy of the one-step UKF is inferior than the two-step UKF since the Kalman filter provides the optimal accuracy. 
Next, a modification of the one-step UKF is presented that recovers the accuracy of the two-step UKF filter in the case of linear systems, that is, it specializes to the classical Kalman filter for linear systems without requiring the second ensemble generation. 
Finally, it is numerically shown that in nonlinear systems with linear output, the modified one-step UKF recovers the accuracy of the two-step UKF filter.




\renewcommand*{\bibfont}{\small}
\printbibliography

\section{APPENDIX}
\subsection{Proof of Proposition \ref{prop:apply_UKF_to_LS}}
\label{proof:2UKF}
\begin{proof}
Note that
, for $i = 1, \ldots, 2 l_x+1,$ 
\begin{align}
    \hat x_{\sigma_i,k+1} 
        &=
            A_k \hat x_{\sigma_i,k} + B_k u_{k}, 
        \nn 
    \\
    \hat y_{\sigma_i,k+1} 
        &=
            C_{k+1} \hat x_{\sigma_i,k+1},
        \nn
\end{align}
and thus
\begin{align}
    X_{k+1|k}
        &=
            A_k X_{k|k} + H(  B_k u_k ),
        \nn
\end{align}
which implies
\begin{align}
    \hat x_{k+1|k}^{\rm UKF} 
        =
            X_{k+1|k} W
        = 
            A_k \hat x + B_k u_k
        = 
            \hat x_{k+1|k}^{\rm KF}.
    \label{eq:x_prior_UKF_is_KF}
\end{align}
%
The perturbed ensemble is thus
\begin{align}
    \tilde X_{k+1|k} 
        &=
            X_{k+1|k} - H(\tilde X_{k+1|k} W)
        \nn \\
        &=
            A_k 
            \matl
                0 &
                \alpha\sqrt{l_x P  } &
                -\alpha\sqrt{l_x P}
            \matr, \nn
\end{align}
%
%
and thus the prior covariance is
\begin{align}
    P_{k+1|k}^{\rm UKF}
        &=
            A_k 
            \matl
                0 &
                \alpha\sqrt{l_x P} &
                -\alpha\sqrt{l_x P}
            \matr
            W_\rmd
            \nn \\ &\quad 
            \cdot \matl
                0 &
                \alpha\sqrt{l_x P} &
                -\alpha\sqrt{l_x P}
            \matr^\rmT 
            A_k^\rmT 
            + Q_k
        \nn \\
        &=
            A_k 
            P
            A_k^\rmT 
            + Q_k
            \nn
        \\
        &=
            P_{k+1|k}^{\rm KF} .
    \label{eq:P_prior_UKF_is_KF}
\end{align}

The second ensemble is
\begin{align}
     X_{k+1|k}'
        =
            \Bigg[ 
                \hat x_{k+1|k}^{\rm KF} \quad 
                &\hat x_{k+1|k}^{\rm KF} + \alpha\sqrt{l_x P_{k+1|k}^{\rm KF}} 
                \nn \\
                &\quad \quad \quad \hat x_{k+1|k}^{\rm KF}  -\alpha\sqrt{l_x P_{k+1|k}^{\rm KF}}
            \Bigg]
        \nn,
\end{align}
and thus 
\begin{align}
    \tilde X_{k+1|k}'
        &=
            \matl
                0 &
                \alpha\sqrt{l_x P_{k+1|k}^{\rm KF}  } &
                -\alpha\sqrt{l_x P_{k+1|k}^{\rm KF}}
            \matr,
    \nn \\
    \tilde Y_{k+1}
        &=
            C_{k+1} \tilde X_{k+1|k}'
    \nn 
\end{align}
Using \eqref{eq:Pz_UKF_FS}, \eqref{eq:Pez_UKF_FS}, it follows that 
\begin{align}
    P_{z_{k+1|k+1}}^{\rm UKF}
        &=
            C_{k+1} \tilde X_{k+1|k}' W_\rmd \tilde X_{k+1|k}^{' \ \rmT} C_{k+1}^\rmT  + R_{k+1}
            \nn
        \\
        &=
            C_{k+1} P_{k+1|k}^{\rm KF} C_{k+1}^\rmT 
            + R_{k+1}
            \nn
        \\
        &=
            P_{z_{k+1|k+1}}^{\rm KF},
        \label{eq:Pz_UKF_is_KF}
        \\
    P_{e,z_{k+1|k}}^{\rm UKF}
        &=
            \tilde X_{k+1|k}' W_\rmd \tilde Y_{k+1}^\rmT
        \nn 
        \\
        &=
            \tilde X_{k+1|k}'  W_\rmd \tilde X_{k+1|k} ^{' \ \rmT} C_{k+1}^\rmT 
        \nn 
        \\
        &=
            P_{k+1|k}^{\rm KF}  C_{k+1}^\rmT 
        \nn \\
        &=
            P_{e,z_{k+1|k}}^{\rm KF} .
    \label{eq:Pez_UKF_is_KF}
\end{align}
Finally, 
\eqref{eq:x_prior_UKF_is_KF}-\eqref{eq:Pez_UKF_is_KF}
imply
\eqref{eq:UKF2_x_eq_KF_x}-\eqref{eq:UKF2P_eq_KFP}.
%
\end{proof}

\subsection{Proof of Proposition \ref{prop:apply_1UKF_to_LS}}
\label{proof:1UKF}
\begin{proof}
Note that
\begin{align}
    X_{k+1|k}
        &=
            A_k X_{k|k} + H(  B_k u_k ),
        \nn \\
    Y_{k+1} 
        &= C_{k+1} X_{k+1|k}. \nn
\end{align}
and thus
\begin{align}
    X_{k+1|k} W
        &=
            A_k \hat x_{k|k} + B_k u_k ,
        \nn \\
    Y_{k+1} W 
        &=
            C_{k+1} A_k \hat x_{k|k} + C_{k+1} B_k u_k   ,
        \nn 
\end{align}
which implies
\begin{align}
    \tilde X_{k+1|k} 
        &=
            A_k X_k 
            - H( A_k \hat x_{k|k} )
        \nn 
        \\
        &=
            A_k 
            \matl
                0 &
                \alpha\sqrt{l_x P_{k|k}^{\rm UKF}  } &
                -\alpha\sqrt{l_x P_{k|k}^{\rm UKF}}
            \matr,
    \label{eq:tildeX_lin_sys}
        \\
    \tilde Y_{k+1} 
        &=
            Y_{k+1} - H(Y_{k+1}W)
        \nn
        \\
        &=
            C_{k+1} \tilde X_{k+1}.
        \label{eq:tildeY_lin_sys}
\end{align}
%

%
The prior covariance is
\begin{align}
    P_{k+1|k}^{\rm UKF}
        &=
            A_k 
            \matl
                0 &
                \alpha\sqrt{l_x P_{k|k}} &
                -\alpha\sqrt{l_x P_{k|k}}
            \matr
            W_\rmd
            \nn \\ &\quad 
            \cdot \matl
                0 &
                \alpha\sqrt{l_x P_{k|k}} &
                -\alpha\sqrt{l_x P_{k|k}}
            \matr^\rmT 
            A_k^\rmT 
            + Q_k
        \nn \\
        &=
            A_k 
            P_{k|k}
            A_k^\rmT 
            + Q_k
            \nn
        \\
        &=
            P_{k+1|k} \nn.
\end{align}
It follows from \eqref{eq:Pz_UKF_1step} and \eqref{eq:Pez_UKF_1step} that
\begin{align}
    P_{z_{k+1|k+1}}^{\rm UKF}
        &=
            C_{k+1} \tilde X_{k+1} W_\rmd \tilde X_{k+1}^\rmT C_{k+1}^\rmT  + R_{k+1}
            \nn
        \\
        &=
            C_{k+1} A_k P_{k|k}  A_k^\rmT C_{k+1}^\rmT + R_{k+1}
            \nn
        \\
        &=
            C_{k+1} (P_{k+1|k} - Q_k ) C_{k+1}^\rmT + R_{k+1}
            \nn
        \\
        &=
            C_{k+1} P_{k+1|k} C_{k+1}^\rmT 
            - C_{k+1} Q_k C_{k+1}^\rmT 
            + R_{k+1}
            \nn
        \\
        &=
            P_{z_{k+1|k+1}} - C_{k+1} Q_k C_{k+1}^\rmT ,
        \label{eq:UKF_Pz}
\end{align}
and
\begin{align}
    P_{e,z_{k+1|k}}^{\rm UKF}
        &=
            \tilde X_{k+1} W_\rmd \tilde Y_{k+1}^\rmT
        \nn 
        \\
        &=
            \tilde X_{k+1} W_\rmd \tilde X_{k+1}^\rmT C_{k+1}^\rmT 
        \nn 
        \\
        &=
            A_k P_{k|k} A_k^\rmT C_{k+1}^\rmT 
        \nn 
        \\
        &=
            P_{k+1|k} C_{k+1}^\rmT - Q_k C_{k+1}^\rmT 
        \nn
        \\
        &=
            P_{e,z_{k+1|k}} - Q_k C_{k+1}^\rmT .
    \label{eq:UKF_Pez}
\end{align}
Since $Q_k \neq 0$ and $C_k \notin \SN(Q_k),$ it follows that 
$P_{z_{k+1|k+1}}^{\rm UKF} \neq P_{z_{k+1|k+1}}$ and 
$P_{e,z_{k+1|k}}^{\rm UKF} \neq P_{e,z_{k+1|k}}$
are missing $C_{k+1} Q_k C_{k+1}^\rmT$ and $Q_k C_{k+1}^\rmT$, respectively, 
, thus implying \eqref{eq:UKF1P_neq_KFP}.
%
%

To prove \eqref{eq:UKF1_K_neq_KF_K}, note that 
\begin{align}
    K_{k+1 } ^{\rm UKF}
        &=
            (P_{e,z_{k+1|k}} - Q_k C_{k+1}^\rmT)
            \nn \\ &\quad 
            \cdot (P_{z_{k+1|k+1}} - C_{k+1} Q_k C_{k+1}^\rmT) ^{-1} ,
        \nn \\
        &\neq 
            K_{k+1}.
        \label{eq:KKFneqKUKF}
\end{align}
Finally, since $K_{k+1}$ minimizes $\tr P_{k+1|k+1}$,
\eqref{eq:KKFneqKUKF} implies \eqref{eq:UKF1P_le_KFP}.
%
\end{proof}
%
\end{document}